\documentclass[aps,prb,twocolumn,showpacs,floatfix,longbibliography]{revtex4-1} 
\usepackage{amsfonts}
\usepackage{graphicx}
\usepackage{amsmath}
\usepackage{times}
\usepackage{amssymb}
\usepackage{color}
\usepackage[colorlinks,bookmarks=false,citecolor=blue,linkcolor=red,urlcolor=blue]{hyperref}

\begin{document}

\title{Quantum phase diagram of spin-$1$ $J_1-J_2$ Heisenberg model on the square lattice: an infinite projected entangled-pair state and density matrix renormalization group study}
\author{R. Haghshenas$^1$}
\email[]{reza.haghshenas@csun.edu}
\author{Wang-Wei Lan$^2$}
\author{Shou-Shu Gong$^3$}
\email{shoushu.gong@buaa.edu.cn}
\author{D. N. Sheng$^1$}
\affiliation{$^1$Department of Physics and Astronomy, California State University, Northridge, California 91330, USA\\ $^2$D\'epartment de Physique, Universit\'e de Sherbrooke, Qu\'ebec, Canada\\ $^3$Department of Physics, Beihang University, Beijing, 100191, China}

\begin{abstract}
We study the spin-$1$ Heisenberg model on the square lattice with the antiferromagnetic nearest-neighbor $J_1$ and the next-nearest-neighbor $J_2$ couplings by using the infinite projected entangled-pair state (iPEPS) ansatz and density matrix renormalization group (DMRG) calculation. The iPEPS simulation, which studies the model directly in the thermodynamic limit, finds a crossing of the ground state from the N\'eel magnetic state to the stripe magnetic state at $J_2/J_1 \simeq 0.549$, showing a direct phase transition. In the finite-size DMRG calculation on the cylinder geometry up to the cylinder width $L_y = 10$, we find that around the same critical point the N\'eel and the Stripe orders are strongly suppressed, which implies the absent of an intermediate phase. Both calculations identify that the stripe order comes with a first-order transition at $J_2/J_1 \simeq 0.549$. Our results indicate that unlike the spin-$1/2$ $J_1-J_2$ square model, quantum fluctuations in the spin-$1$ model may not be strong enough to stabilize an intermediate non-magnetic phase.
\end{abstract}

\maketitle

%%%%%%%%%%%%%%%%%%%%%%%%%%%%%%%%%%%%%%%%%%%%% 
\section{Introduction}
Frustrated magnetic systems play a key role in understanding the exotic phases of matter~\cite{Balents:2010,savary:2016}. As we know, frustration can enhance quantum fluctuations by posing incompatibility on the local interaction energy to be simultaneously satisfied, which may destroy magnetic long-range order and lead the systems into novel quantum phases, such as valence-bond solid (VBS)~\cite{Affleck:1987, read:1989, read:1990} and quantum spin liquid~\cite{Anderson:1973, Wen:1990, Read:1991}. In addition, quantum phase transitions between such phases may defy the Ginzburg-Landau theory---referred to the deconfined quantum criticality~\cite{Senthil:2004}---which makes the physics of frustrated quantum magnetism a fascinating subject both in theoretical and experimental senses. Among the various frustrated antiferromagnets, the spin-$S$ $J_1-J_2$ square Heisenberg models~\cite{IOFFE:1988, Zhitomirsky:1996, Xu:1990, Bruder:1992} are well-known examples and have stimulated extensive theoretical studies over the last two decades. The competing interactions in such systems may stabilize a non-magnetic intermediate phase if quantum fluctuations are not strongly suppressed~\cite{IOFFE:1988, Zhitomirsky:1996, Xu:1990, Bruder:1992}. One of the successful examples is the spin-$1/2$ $J_1-J_2$ model, in which a non-magnetic intermediate phase has been identified by using different methods although the nature of the phase is still controversial~\cite{Jiang:2012, Hu:2013, Wang:2013, Gong:2014, Morita:2015, Wang:2017, haghshenas:2017}.

Meanwhile, the frustrated spin-$1$ square Heisenberg models (SHM) are also typical for studying frustrated magnetism. Recently, the studies on the magnetism of the iron-based superconductors have drawn extensive interests in investigating the novel quantum phases, particularly the non-magnetic phase with lattice nematic order, in different spin-$1$ SHMs~\cite{Xu:2008, Qimiao:2008, Fang:2008, Mazin:2014, Yu:2015, wang:2015, lai:2016, wangzt:2016, Zhu:2016, Hu:2016, Gong:2017, Niesen:2017, Hu:2018}. Among the various models, the spin-$1$ $J_1 - J_2$ SHM is probably the most fundamental model, which is defined as
\begin{equation*}
H=J_1 \sum_{\langle i,j\rangle}\textbf{S}_{i}\cdot\textbf{S}_{j}
+J_2 \sum_{\langle\langle i,j\rangle\rangle}\textbf{S}_{i}\cdot\textbf{S}_{j},
\end{equation*}
where $\langle i,j\rangle$ and $\langle \langle i,j\rangle \rangle$ denote the nearest-neighbor and the next-nearest-neighbor pairs, and $J_1$ and $J_2$ are both antiferromagnetic (AFM) couplings. We set $J_1 = 1$ as the energy scale. Classically, this model in the large $S$ limit has a N\'eel and a stripe AFM phase separated at $J_2 = 0.5$. After considering quantum fluctuations, the early studies based on the modified spin-wave theory~\cite{Xu:1990, Gochev:1995} and Schwinger-Boson mean-field theory\cite{Mila:1991} predicted that quantum fluctuations for the systems with spin magnitude $S > 0.7$ are not strong enough to stabilize a non-magnetic intermediate phase. Thus, mean-field results suggested a direct phase transition from the N\'eel to the stripe AFM phase for the spin-$1$ model. This result was later confirmed by the coupled cluster method, which found a first-order transition between the N\'eel and stripe phase at $J_2 \simeq 0.55$~\cite{Bishop:2008}. Interestingly, the recent density matrix renormalization group (DMRG) study~\cite{Jiang:2009} challenged this result: it predicted a non-magnetic phase in the small intermediate region for $0.525 \lesssim J_2 \lesssim 0.555$, and suggested that this non-magnetic phase might be continuously connected to the limit of the decoupled Haldane spin chains~\cite{Haldane:1983} by tuning the spacial anisotropic couplings $J_{1x}$ and $J_{1y}$. Such a non-magnetic phase in spin-$1$ model is quite interesting not only because it goes beyond the physics in the mean-field description, but also because it might be related to the nematic non-magnetic phase in the iron-based superconductor material FeSe~\cite{wang:2015}.

In this article, our main goal is to reexamine the phase diagram of the spin-$1$ $J_1 - J_2$ SHM based on the variational tensor-network ansatz and DMRG simulation~\cite{chen:2018}. In previous studies, while the mean-field calculation may not fully consider quantum fluctuations~\cite{Xu:1990, Gochev:1995, Mila:1991}, the DMRG simulation may have difficulty to pin down the intermediate region as a non-magnetic phase due to the finite-size effects~\cite{Jiang:2009}. To this end, we use the state-of-the-art numerical methods to systematically study the model. We use the $U(1)$-symmetric infinite projected entangled-pair state (iPEPS) ansatz \cite{Jordan:2008, Murg:2009} and the $SU(2)$-symmetric finite-size DMRG to study the system from different limits: the iPEPS is directly applied in the thermodynamic limit, significantly diminishing possible finite-size effects; $SU(2)$-symmetric DMRG obtains accurate results on finite-size system. In the iPEPS ansatz, the only control parameter is the so-called bond dimension $D$ which controls entanglement in the system. To simulate highly entangled states (larger bond dimensions), we implement $U(1)$ symmetry into the iPEPS ansatz~\cite{Singh:2010}. We expect that a close comparison of these different approaches could substantially improve our understanding of the intermediate regime.

In our iPEPS simulation, we use a hysteresis effect to accurately determine the quantum phase transition and its nature~\cite{Corboz:2013}. We initialize the iPEPS ansatz with different types of wavefunctions to find the lowest variational ground-state energy. It is found that either the N\'eel or the stripe state provides the lowest iPEPS energy throughout the coupling parameter range. The energy of the N\'eel and the stripe states cross each other at the critical point $J_2 \simeq 0.549$, where the magnetic order parameters (defined later) remain non-zero in the $D \rightarrow \infty $ limit, suggesting a first-order phase transition between the two magnetic order phases. We also observe that the correlation length is unlikely to show a divergent behavior around the transition point, further supporting a first-order transition. In our DMRG calculation with the improved system size for size-scaling analysis, we find that the N\'eel order could persist to $J_2 \simeq 0.545$ and the stripe order grows up sharply at $J_2 \simeq 0.55$, showing a very narrow non-magnetic regime, which is quite smaller than the previous DMRG result $\sim 0.03$~\cite{Jiang:2009}. The fast-shrinking intermediate regime, which is observed in the finite-size scaling with increased system size in the DMRG results, may suggest the vanishing non-magnetic phase. The non-magnetic phase with a strong lattice nematicity in spin-$1$ models, which has been proposed for the nematic paramagnetic phase of FeSe~\cite{wang:2015}, might be stabilized by considering other compting and/or frustrating interactions.

The paper is organized as follows. We briefly discuss the numerical methods and define the order parameters used in this paper in Sec.~\ref{Sec:METHOD}. Our main numerical results are presented in Sec.~\ref{sec:Numericalresults}. In Sec.~\ref{sec:iPEPSresults}, we compare the variational ground-state energy obtained by the numerical methods and discuss the iPEPS results by studying the variational energy of competitive ordered states and behavior of the spin correlation length. In Sec.~\ref{sec:DMRGresults}, we provide a systematic study of the nematic and magnetic order parameters and also discuss the behavior of the gap by using DMRG. Finally, we summarize our findings in Sec.~\ref{Sec:CONCLUSION}.

%%%%%%%%%%%%%%%%%%%%%%%%%
\section{METHODS}
\label{Sec:METHOD}
%%%%%%%%%%%%%%%%%%%%%%%%%%%%%%%%%%Fig. 2%%%%%%%%%%%%%%%%%%%%%%%%%%%%%%%%%%%%%%%%%%
\begin{figure}
\begin{center}
\includegraphics[width=1.0 \linewidth]{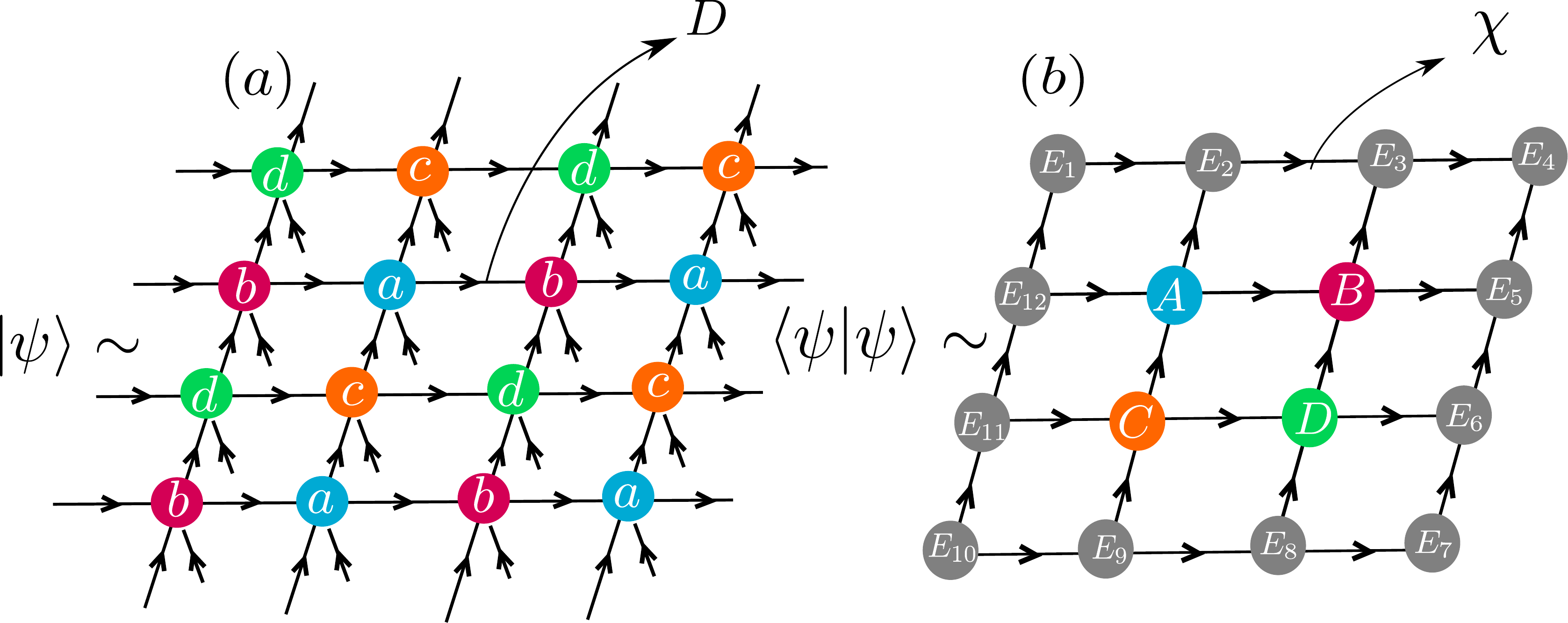} 
  \caption{(Color online) Tensor-network representation of a $U(1)$-symmetric iPEPS $|\psi\rangle$. $(a)$ The iPEPS $|\psi\rangle$ is made of the $U(1)$-invariant tensors $\{a, b, c, d\}$ periodically repeated through the infinite square lattice. $(b)$ The scalar product $\langle\psi|\psi\rangle$ is calculated by using the so-called environment tensors $\{E_{1},\cdots,E_{12}\}$ obtained by CTMRG approach. The bond dimensions $D, \chi$ controls accuracy of the iPEPS anstaz.}
  \label{fig:ipeps}
\end{center}
\end{figure}
 %%%%%%%%%%%%%%%%%%%%%%%%%%%%%%%%%%%%%%%%%%%%%%%%%%%%%%%%%%%%%%%%%%%%%%%%5

\subsection{iPEPS anstaz}

The iPEPS ansatz provides an efficient variational method to approximate the ground-state wave functions of the two-dimensional spin systems in the thermodynamic limit~\cite{Verstraete:2008}. The iPEPS is made of some building-block tensors which are periodically repeated through the infinite 2D lattice. The tensors are connected to each other by the so-called virtual bonds to construct a geometrical pattern (usually) similar to the 2D lattice. The main idea is to variationally minimize the expectation value of energy with respect to the tensors (variational parameters) to eventually obtain an approximation of the ground state. The bond dimension of the virtual bonds denoted by $D$ determines the number of the variational parameters, hence, controlling the accuracy of the iPEPS ansatz. It also represents the amount of entanglement, so that by increasing it even `highly entangled states' could be accurately approximated~\cite{Eisert:2013}.

In this paper, as shown in Fig.~\ref{fig:ipeps}(a), we use (up to) a $2 \times 2$ unit cell iPEPS constructed from five-rank independent tensors $\{a, b, c, d\}$. We exploit $U(1)$ symmetry, making all tensors $\{a, b, c, d\}$ to be $U(1)$ invariant, to reach the larger bond dimensions up to $D\sim 9$~\cite{Bauer:2011}. We perform the optimization procedure by using imaginary-time evolution in the class of the iPEPS~\cite{Corboz:2010}
\begin{equation*}
 |\psi_{i+1} \rangle=e^{-\tau H} |\psi_{i} \rangle,
\end{equation*}
where $|\psi_{i+1}\rangle$ at each step $i$ is represented by an iPEPS---$\tau$ stand for imaginary time. A first-order Trotter-Suzuki decomposition~\cite{Suzuki:1990} is used to efficiently represent the imaginary time-evolution operator $e^{-\tau H}$. We also use the so-called full-update scheme\cite{haghshenas:2017, Phien:2015} to truncate the bond dimension: at each step, imaginary time-evolution operator increases the bond dimension of virtual bonds, so a truncation procedure is needed to prevent from the exponential growth of the parameters. In the case of the spin-$1$ $J_1 - J_2$ SHM, we find that the computationally cheaper scheme, i.e., the simple update~\cite{Jiang:2008,Gu:2008,Corboz:2010:April}, does not give us accurate results, but, its output could be used as a good initial state for the full-update scheme. 

A corner transfer matrix renormalization group (CTMRG) approach\cite{Nishino:1997,Corboz:2014, Huang:2012} is used to evaluate the expectation values of observables and also to obtain the so-called environment tensors (needed within optimization procedure). The accuracy of CTMRG is controlled by the `boundary' bond dimension of the environment tensors denoted by $\chi$ (see Fig.~\ref{fig:ipeps}(b)). We always take $\chi$ large enough to diminish the error due to the environment approximation: for the largest bond dimension $D=9$, it approximately requires $\chi \sim 100$ to make the relative error of the ground-state energy negligible.
 
In order to recognize magnetically ordered phases, we calculate magnetic order parameter defined by
\begin{align*}
m=\frac{1}{4}  (|\langle  \textbf{S}_{a} \rangle| + |\langle \textbf{S}_{b}\rangle|  + |\langle  \textbf{S}_{c}\rangle| +  |\langle \textbf{S}_{d}  \rangle| ), 
\end{align*}
where operator $ \textbf{S}_{a} $ is acting on tensor $a$ (analogous for other operators). Since the magnetically ordered states break $SU(2)$ symmetry, so we expect to obtain a non-zero value of $m$ in the $D \rightarrow \infty$ limit. 

In addition, we use the texture of the local bond $J_1$ energy to detect the lattice symmetry breaking, as defined by $\Delta T_{x(y)}=\max(E_{x(y)})-\min(E_{x(y)})$, and $\Delta T_{x-y} = \max(E_{y})-\min(E_{x})$. The order of taking maximum and minimum is chosen to enlarge the order parameters; although, it does not significantly affect the results. The symbols $E_{x}$ and $E_{y}$ stand for the local bond energy in the unit cell for the horizontal and vertical directions---note for each virtual bond, we might obtain a different value of $E_{x, y}$. The order parameters $\Delta T_{x-y}$ and $\Delta T_{x}$ respectively detect rotational and translational lattice symmetry breaking. Similarly, a finite value of $\Delta T_{x-y, x}$ in the $D\rightarrow \infty$ limit implies a lattice symmetry breaking. By using the order parameters $\{m, \Delta T_{x-y, x}\}$, we could distinguish between different types of ordered states, such as N\'{e}el, stripe, VBS and Haldane.

In order to estimate the variational energy and the order parameters, we use a polynomial and linear fit with bond dimension $1/D$, respectively. Our intuition to using such extrapolations is that it provides an accurate estimation of these quntities at point $J_2=0$.  The results could be compared to that of quantum Monte-Carlo method \cite{Matsumoto:2001}, as at this point, there is no sign problem. For example, the relative error of our estimation of the magnetic order parameter is of order $\Delta m= \frac{m_{D \rightarrow \infty}-m_{\text{MC}}}{m_{\text{MC}}}<10^{-2}$, see also Ref.~\onlinecite{Niesen:2017} for similar results.

Finally, we utilize the spin-spin correlation function $C^{s}(r)$ and corresponding correlation length $\xi^{s}$ to distinguish a quantum critical point (phase) from the ordered phases. They are defined by
\begin{align*}
&C^{s}(r)=\langle \textbf{S}_{(x, y)} \cdot \textbf{S}_{(x+r, y)} \rangle  -  \langle \textbf{S}_{(x, y)}  \rangle^{2}, \\
&\log(C^{s}(r))=(\frac{-1}{\xi^{s}})r+const \quad r \gg 1.
\end{align*}
Usually, a finite bond dimension $D$ (usually) induces a finite correlation length\cite{Evenbly:2011, Verstraete:2006}, thus, to determine the true nature of the phases, we need to study the correlation length in the large-$D$ limit. For a quantum critical point, by increasing $D$, one expects the correlation length to grow rapidly. On the other hand, for the ordered phases such as N\'eel phase, it tends to a finite value in the large-$D$ limit.

\subsection{DMRG method}

We implement DMRG~\cite{White:1992} for studying finite-size system. A rectangular cylinder (RC) geometry is used in our calculation, which has the periodic boundary conditions in the $y$ direction and the open boundary conditions in the $x$ direction. We denote the cylinder as RC$L_y$-$L_x$, where $L_y$ and $L_x$ are the numbers of sites along the $y$- and $x$-directions, respectively. In order to obtain accurate results on wide cylinder with $L_y$ up to $10$, we use $SU(2)$-symmetric DMRG~\cite{Ian:2002}  by keeping as many as about $20000$ $U(1)$-equivalent states ($6000$ $SU(2)$ optimal states). The truncation error is less than $1 \times 10^{-5}$.

In order to detect the magnetic orders, we calculate the magnetic order parameter $m^2({\bf q}) = \frac{1}{N_s^2} \sum_{i,j} \langle {\bf S}_i \cdot {\bf S}_j \rangle e^{i {\bf q} \cdot ({\bf r}_i - {\bf r}_j)}$, where $N_s$ is the summed total site number. For the N\'eel and the stripe AFM order, the order parameter $m^2({\bf q})$ shows the peak at ${\bf q} = (\pi,\pi)$ and ${\bf q} = (0,\pi)/(\pi,0)$, which are denoted as $m^2(\pi,\pi)$ and $m^2(0,\pi)/(\pi,0)$. For the stripe order, since our cylinder geometry breaks lattice symmetry, DMRG calculation selects the momentum at ${\bf q} = (0, \pi)$. To obtain the order parameters with reduced boundary effects, we use the spin correlation functions of the middle $L_y \times L_y$ sites on the RC$L_y$-$2L_y$ cylinder.

We also define the bond nematic order $\sigma_1$ as the difference between the horizontal and vertical $J_1$ bond energy, namely, $\sigma_1 = \langle {\bf S}_i \cdot {\bf S}_{i+\hat{x}} \rangle - \langle {\bf S}_i \cdot {\bf S}_{i+\hat{y}} \rangle$ ($i$ could be an any lattice site in the bulk of cylinder as the translational symmetry shown below, $\hat{x}$ and $\hat{y}$ are the unit vectors along the $x$ and $y$ directions, respectively) to study the possible lattice rotational symmetry breaking.

\begin{figure}
\includegraphics[width = 1.0\linewidth]{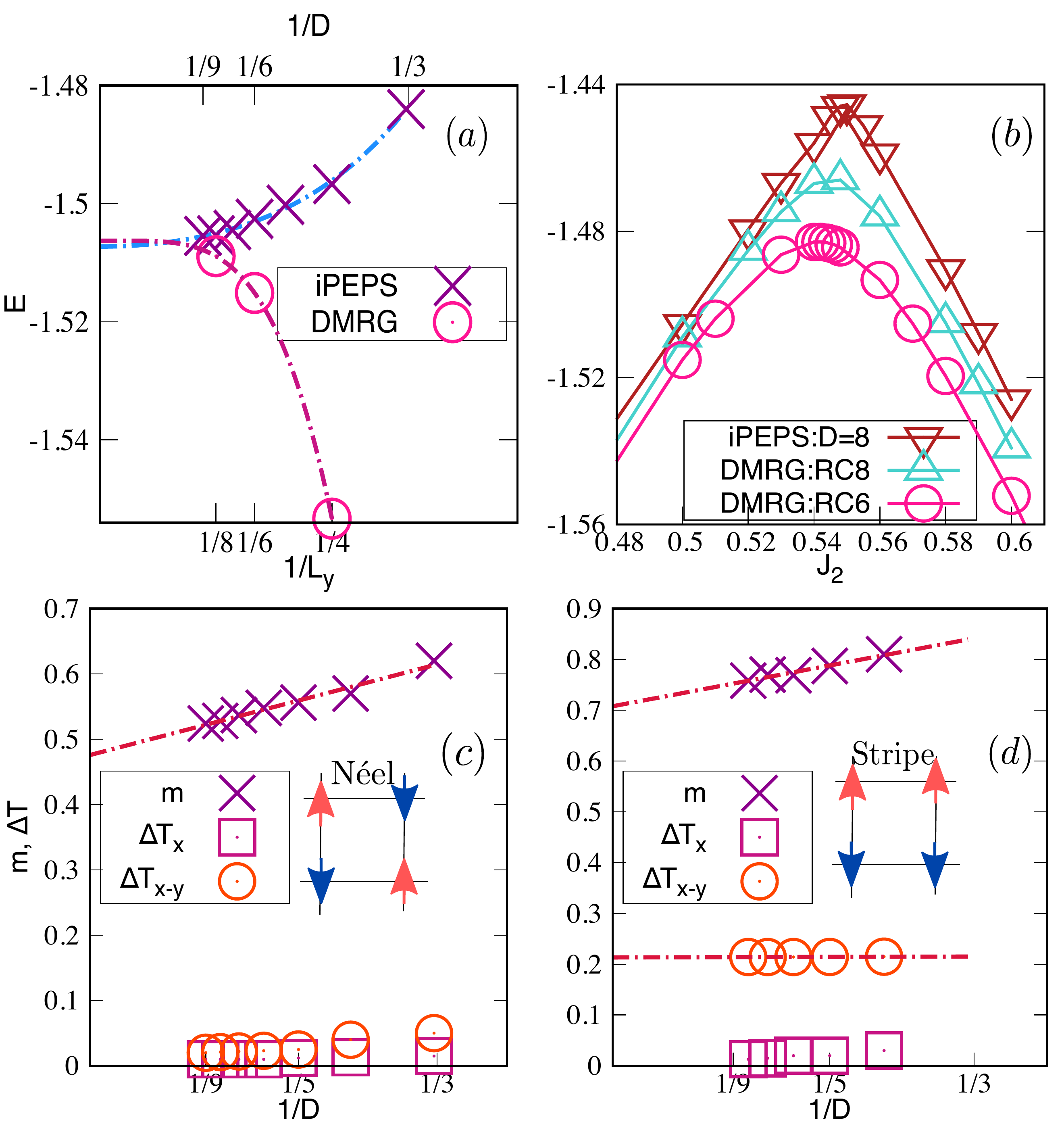}
\caption{ (Color online) Comparing the ground-state energy obtained by iPEPS and DMRG. (a) The ground-state energy as a function of $1/D$ ($1/L_y$) at $J_{2}=0.5$ for iPEPS (DMRG). The DMRG data are for $L_y = 4, 6, 8$. The dashed lines represent polynomial fit up to the fourth order. (b) The ground-state energy as a function of $J_2$. The sharp peak around $J_2\sim 0.55$ suggests a first-order quantum phase transition. The order parameters $\{m, \Delta T_{x}, \Delta T_{x-y}\}$ as a function of $1/D$ for (c) $J_{2}=0.5$ and (d) $J_{2}=0.6$. They are respectively compatible with a N\'eel and stripe AFM order. The insets (graphical figures) show the pattern of the magnetization.}
\label{fig:energy}
\end{figure}
%%%%%%
%%%%%% 
\begin{figure}
\includegraphics[width = 1.0\linewidth]{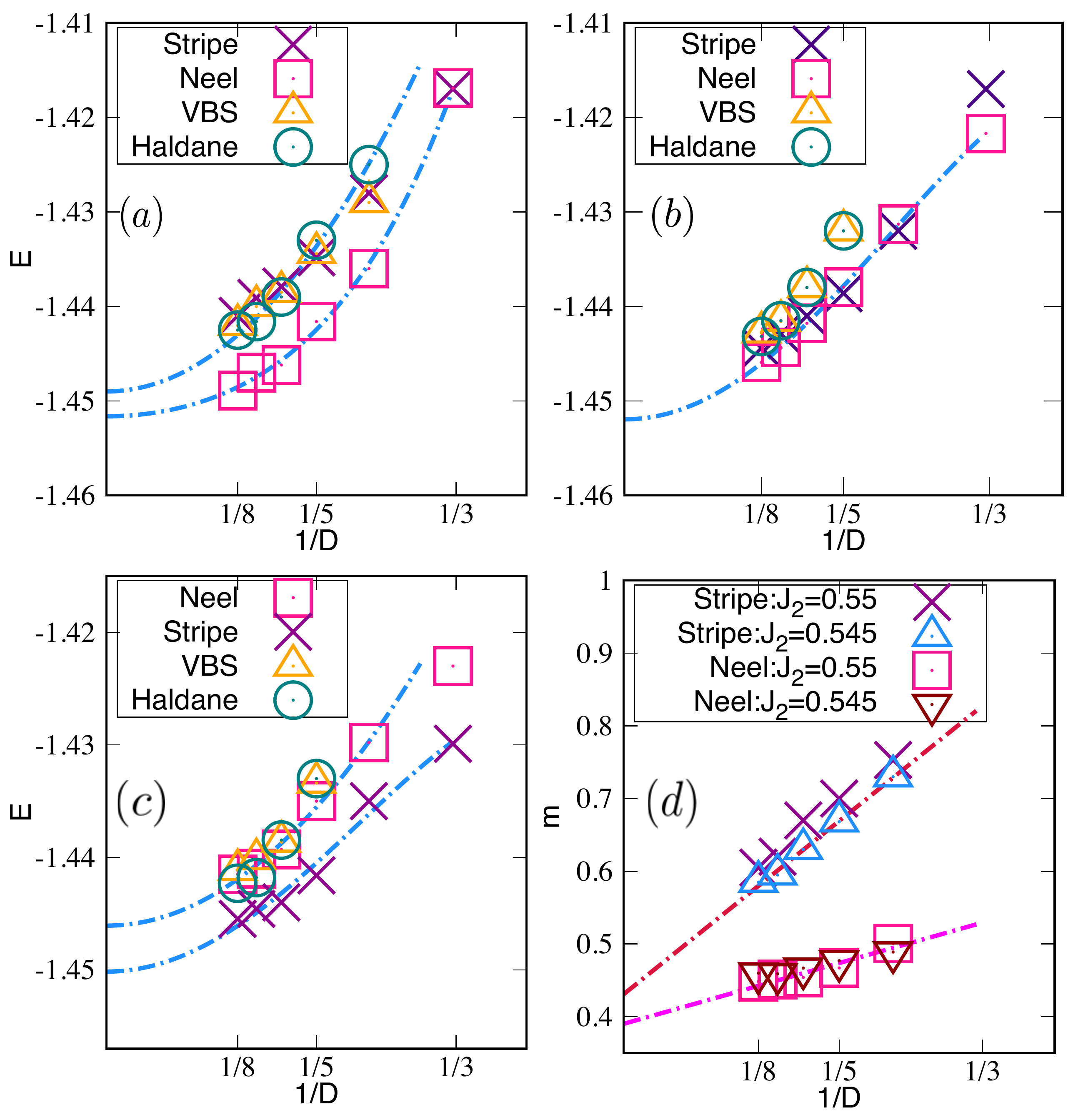}
\caption{(Color online) A hysteresis analysis. (a)-(c) are the iPEPS ground-state energy initialized using different states of the N\'{e}el, stripe, VBS and Haldane at $J_2 = 0.545, 0.548, 0.55$. (d) The magnetic order parameter $m$ for $J_2 = 0.545, 0.55$. A non-zero magnetic order parameter in the large-$D$ limit implies a first-order quantum phase transition. In stripe (N\'{e}el) phase, due to hysteresis effect, the order parameter $m$ with a N\'{e}e (stripe) pattern remains metastable.}
\label{fig:Bond-energy}
\end{figure}
%%%%%%

%%%%%%%%%%%%%%%%%%%%%%%%%%%%%%%%%%%%%%%%%%%%%%%%%%%%
\section{Numerical results} 
\label{sec:Numericalresults}
\subsection{iPEPS results}
\label{sec:iPEPSresults}

We start by comparing the ground-state energy obtained by the iPEPS and finite-size DMRG as shown in Figs.~\ref{fig:energy}(a, b). The benchmark results agree quite well at the highly frustrated point $J_2 = 0.5$: the extrapolated values from a polynomial fit reveal that $E^{D \rightarrow \infty}_{\text{iPEPS}} \simeq -1.507$ and $E^{L_y \rightarrow \infty}_{\text{DMRG}}  \simeq -1.506$. We find that the best fitting curves for the DMRG and the iPEPS energy are obtained by function $f(x)=a+\frac{b}{x^{2}}+\frac{c}{X^{4}}$, where $x=\frac{1}{D}, \frac{1}{L_{y}}$. With growing $J_2$, we observe that the energy monotonically increases to $J_{2} \sim 0.55$ through a sharp peak and then it starts to decrease, which may suggest a first-order quantum phase transition.

In Figs.~\ref{fig:energy}(c, d), we present the iPEPS results of the order parameters $\{m, \Delta T_{x}, \Delta T_{x-y}\}$ to identify the nature of the ground state at $J_{2}=0.5, 0.6$---since $\Delta T_{y}$ behaves similar to $\Delta T_{x}$ in our results, we only show the latter. For $J_2=0.5$, we obtain a finite value of $m \sim 0.48$ in the large-$D$ limit, and its configuration is compatible with an AFM N\'{e}el state (see the inset, the graphical figure). The order parameters $\Delta T_{x}, \Delta T_{x-y}$ are strongly suppressed by increasing $D$, which agrees with a N\'eel state. The N\'{e}el state is strongly established at this point: even if we initialize the iPEPS with a random state or a stripe state, the outcome of the iPEPS simulation always gives a N\'{e}el state. Similarly, a stripe AFM state is established at $J_{2}=0.6$ as observed by the pattern of the local magnetic order (see the inset in Fig.~\ref{fig:energy}(d)), the non-zero $\Delta T_{x-y} \simeq 0.2$, and the vanished $\Delta T_{x}$. Therefore, we establish the N\'eel phase and the stripe phase in the small and large $J_2$ side, respectively.
 
Next, we focus on the intermediate regime for $J_{2} \sim 0.5-0.6$. We use the hysteresis analysis~\cite{note:hysteresis} (see Ref.~\onlinecite{Corboz:2013}) to study phase transitions in this regime. The main idea is to initialize the iPEPS with different possible states and find whether the energies cross each other. The crossing of energy is considered as an evidence of a quantum phase transition. Particularly, if in the vicinity of this crossing the order parameters (strongly) remain non-zero, it indicates a first-order phase transition. In addition to the N\'eel and stripe states, we also use the columnar VBS and the Haldane state---both of them could be competitive candidates for a paramagnetic intermediate phase \cite{Sadrzadeh:2016, haghshenas:2017}. The Haldane state can be obtained by simply setting $J_{2} = 0$ and  $J_{1y} = 0$ (the nearest-neighbor coupling along the $y$-direction), which results in a set of decoupled 1D Haldane chains. The non-magnetic intermediate phase found in the previous DMRG calculation was proposed to be continuously connected to the Haldane phase~\cite{Jiang:2009}. The columnar VBS state consists of staggered singlet bonds (connecting all two adjacent spins) along either the $x$ or the $y$ directions.

In Figs.~\ref{fig:Bond-energy}(a-c), we have plotted the iPEPS ground-state energy initialized with different states at $J_2 = 0.545, 0.548, 0.55$. For $J_2 = 0.545$, the N\'eel state gives the lowest energy. By increasing $J_{2}$, the energies get closer and, at $J_2=0.548$, it is found that the N\'{e}el and stripe states have almost the same energy---the N\'{e}el state still has slightly lower energy in the large-$D$ limit. Finally, at $J_{2}=0.55$, the energy of the stripe state obviously becomes the lowest one. Our iPEPS results show that the energies of the N\'{e}el and stripe state cross each other at $J_{2} \sim 0.549$. In this region, the energy of the Haldane state is always lower than the columnar VBS in the large-$D$ limit; however, the columnar VBS and the Haldane state never have the lower energy than the magnetically ordered states. 
%%%%%%%%%%%%%%%%%%%%%%%%%%%%%%%%%%%%%%%%%%%%
\begin{figure}[b]
\includegraphics[width = 1.0\linewidth]{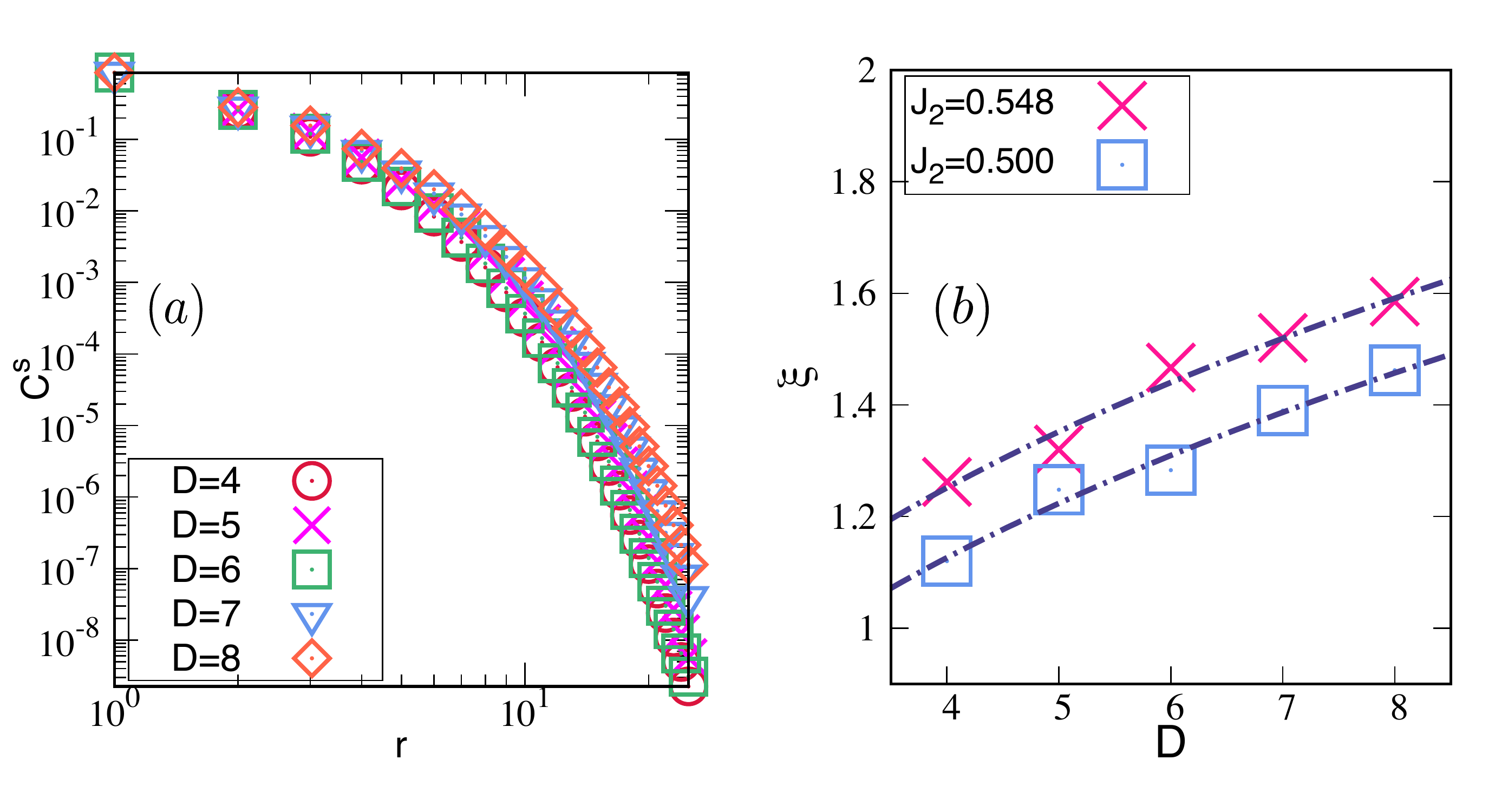}
\caption{(Color online) Spin correlation function close to the quantum phase transition point. (a) Log-log plot of the spin-spin correlation function versus the site distance $r$. (b) The correlation lengths as a function of bond dimension $D$ for $J_{2} = 0.5, 0.548$. The dashed lines are power-law fits $\xi^{s} \sim D^{\alpha}$.}
\label{fig:Corr}
\end{figure}
%%%%%%%%%%%%%%%%%%%%%%%%%%%%%%%%%%%%%%%%%%%%%

We study the magnetic order parameter $m$ to investigate the type of the quantum phase transition where the energies cross. As shown in Fig.~\ref{fig:Bond-energy}(d), $m$ shows the linear decreasing as a function of $1/D$: we obtain quite large values of the magnetization in the region $J_2 \sim 0.545-0.55$. For both $J_2 = 0.545, 0.55$, the N\'eel and the stripe orders show $m = 0.39, 0.43$ in the $D\rightarrow \infty$ limit. The non-zero magnetization through the quantum phase transition implies a first-order transition. Note, due to the hysteresis effect, in the vicinity of a first-order quantum phase transition, the order parameter $m$ with a N\'eel (stripe) pattern remain non-zero even in the stripe (N\'eel) phase\cite{note:hysteresis}.

Furthermore, we study the correlation function as another probe to investigate the quantum phase transition. In Fig.~\ref{fig:Corr}(a), we demonstrate the log-log plot of the spin-spin correlation function $C(r)$ at $J_2=0.548$ for different values of bond dimension $D$. It seems that $C(r)$ would have an exponential fall-off, as it weakly depends on the bond dimension $D$. To get more insight, we compare the behavior of the correlation lengths at $J_2 = 0.549$, very close to the transition point, with the one deep inside the N\'eel phase (at the point $J_2=0.50$). For a continuous quantum phase transition, one expects $\xi^{s}$ to sharply grow as the system gets close to the critical point. We plot $\xi^{s}$ as a function of bond dimension $D$ for $J_2=0.50, 0.549$ in Fig.~\ref{fig:Corr}(b). The correlation length illustrates almost the same behavior: it similarly grows as $\xi^{s} \sim D^{0.35}$, which seems not to support a divergent correlation length at the transition point but could be consistent with a first-order transition.
\begin{figure}
\includegraphics[width = 0.90\linewidth]{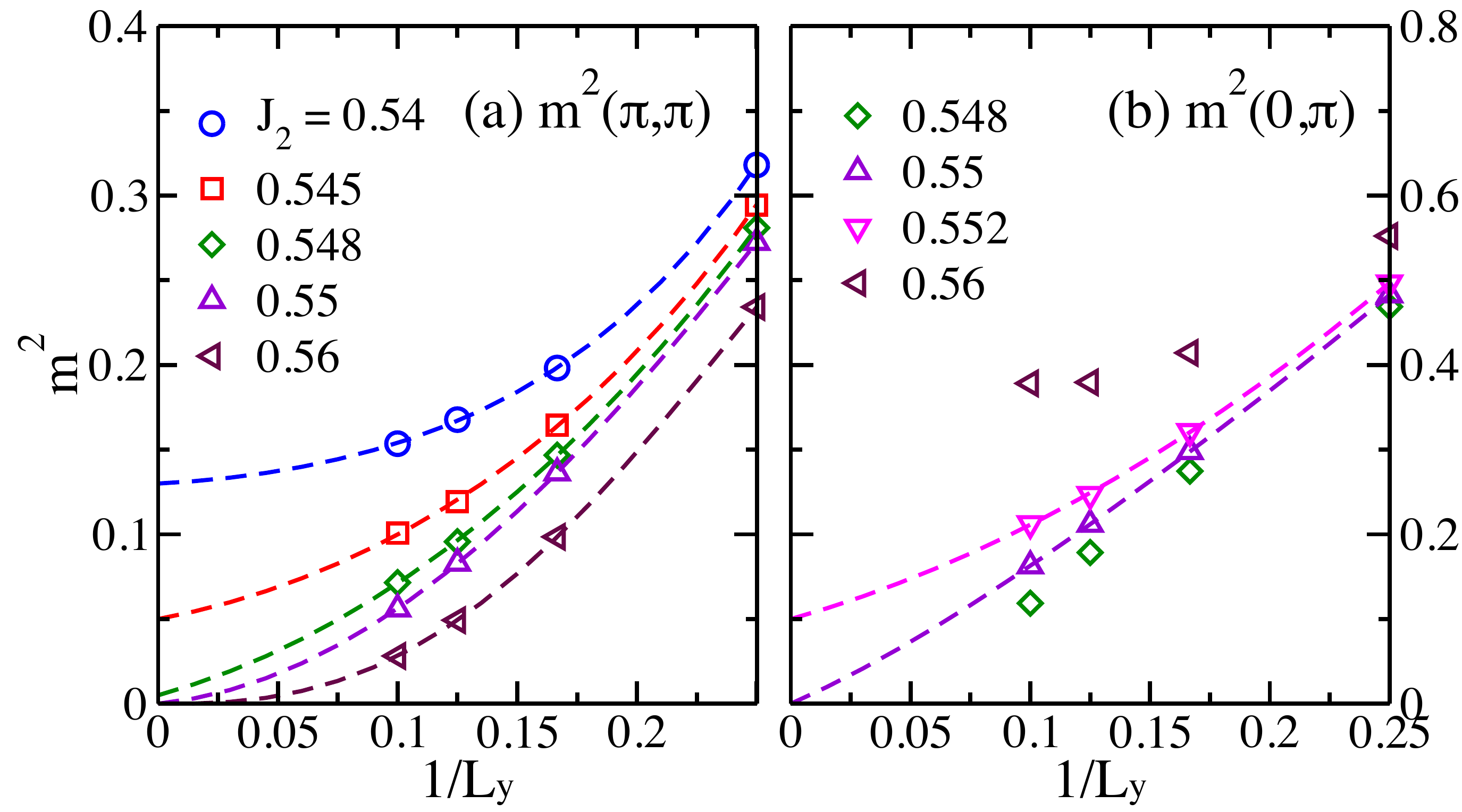}
\includegraphics[width = 0.90\linewidth]{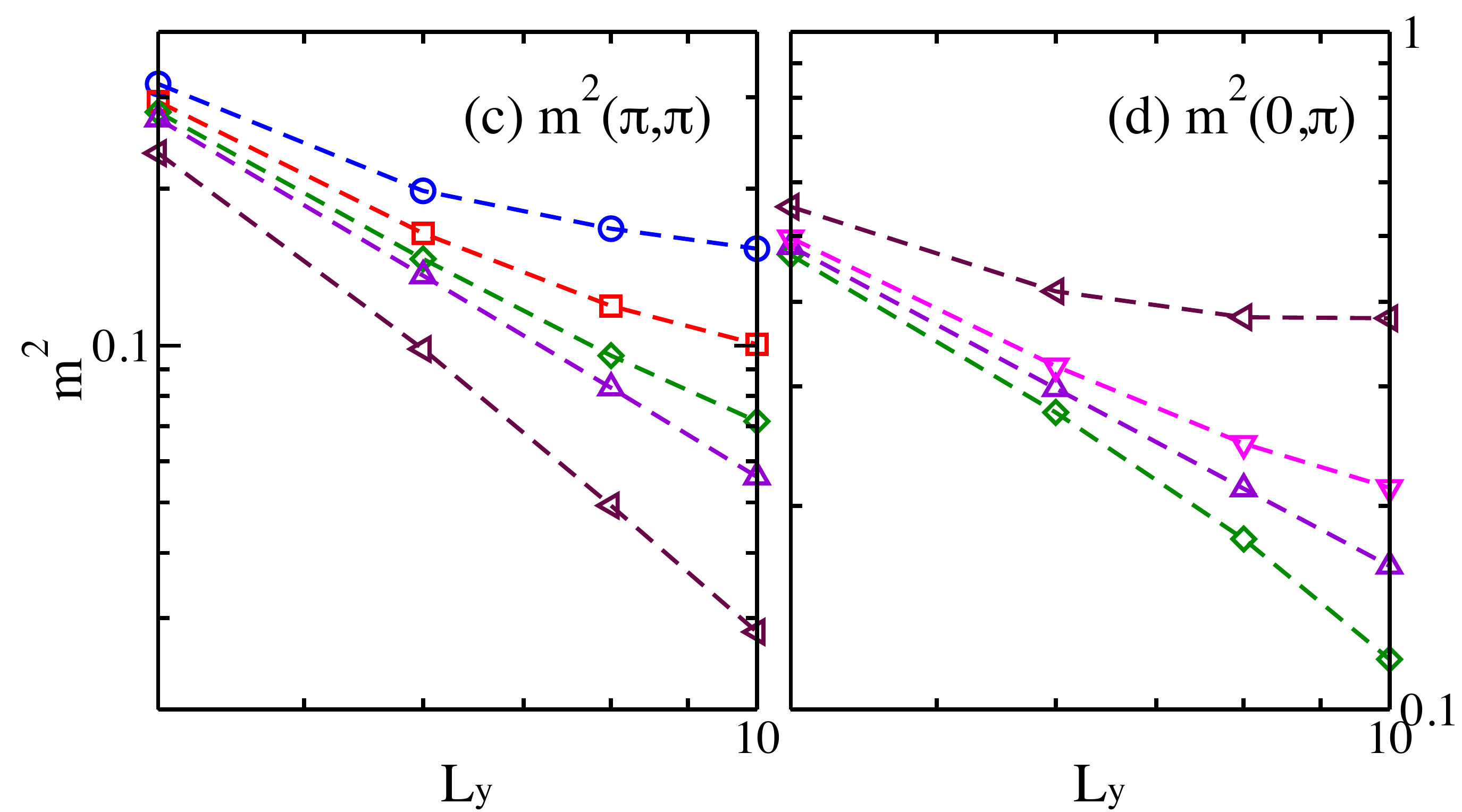}
\caption{(Color online) Finite-size scaling of magnetic order parameters on the RC$L_y$-$2L_y$ cylinders with $L_y = 4, 6, 8, 10$. (a) and (b) are the N\'{e}el and the stripe order parameters $m^2(\pi,\pi)$ and $m^2(0,\pi)$ versus $1/L_y$, respectively. The dashed lines are the polynomial fits up to fourth orders. (c) and (d) are the corresponding log-log plots versus $L_y$, where the dashed lines are guides to the eye.}
\label{fig:j1j2}
\end{figure}
%%%%%%

%%%%%% 
\subsection{DMRG results}
\label{sec:DMRGresults}

Next, we demonstrate our DMRG results on the finite-$L_y$ cylinder system. First of all, we show the magnetic order parameters $m^2(\pi,\pi)$ and $m^2(0,\pi)$ with $L_y$ from $4$ to $10$ in Figs.~\ref{fig:j1j2}(a, b). Through appropriate finite-size scaling, we find that the N\'eel order $m^2(\pi,\pi)$ could persist to $J_2 \simeq 0.545$. For $J_2 > 0.55$, the stripe order $m^2(0,\pi)$ develops very fast with growing $J_2$, as shown by $J_2 = 0.552$ in Fig.~\ref{fig:j1j2}(b). Compared with the previous DMRG results based on the size-scaling up to the $L_y = 8$ torus~\cite{Jiang:2009}, our analysis up to $L_y = 10$ cylinder suggests a much smaller regime $0.545 < J_2 < 0.55$ for a possible intermediate phase. The log-log plots of $m^2$ versus $L_y$ in Figs.~\ref{fig:j1j2}(c, d) also agree with the transition between different orders at $J_2 \simeq 0.55$, where the two magnetic orders change their behaviors dramatically. At $J_2 = 0.55$, both magnetic order parameters seem to follow a critical behavior. Such a critical-like behavior of order parameters could be consistent with a continuous phase transition at $J_2 \simeq 0.55$. Here, we remark that the system size in our DMRG calculation is too small for such a critical analyses. And previous studies~\cite{Bishop:2008} and our results have already shown that the transition at $J_2 \simeq 0.55$ is a first-order transition.

%%%%%% 
\begin{figure}[b]
\includegraphics[width = 0.7\linewidth]{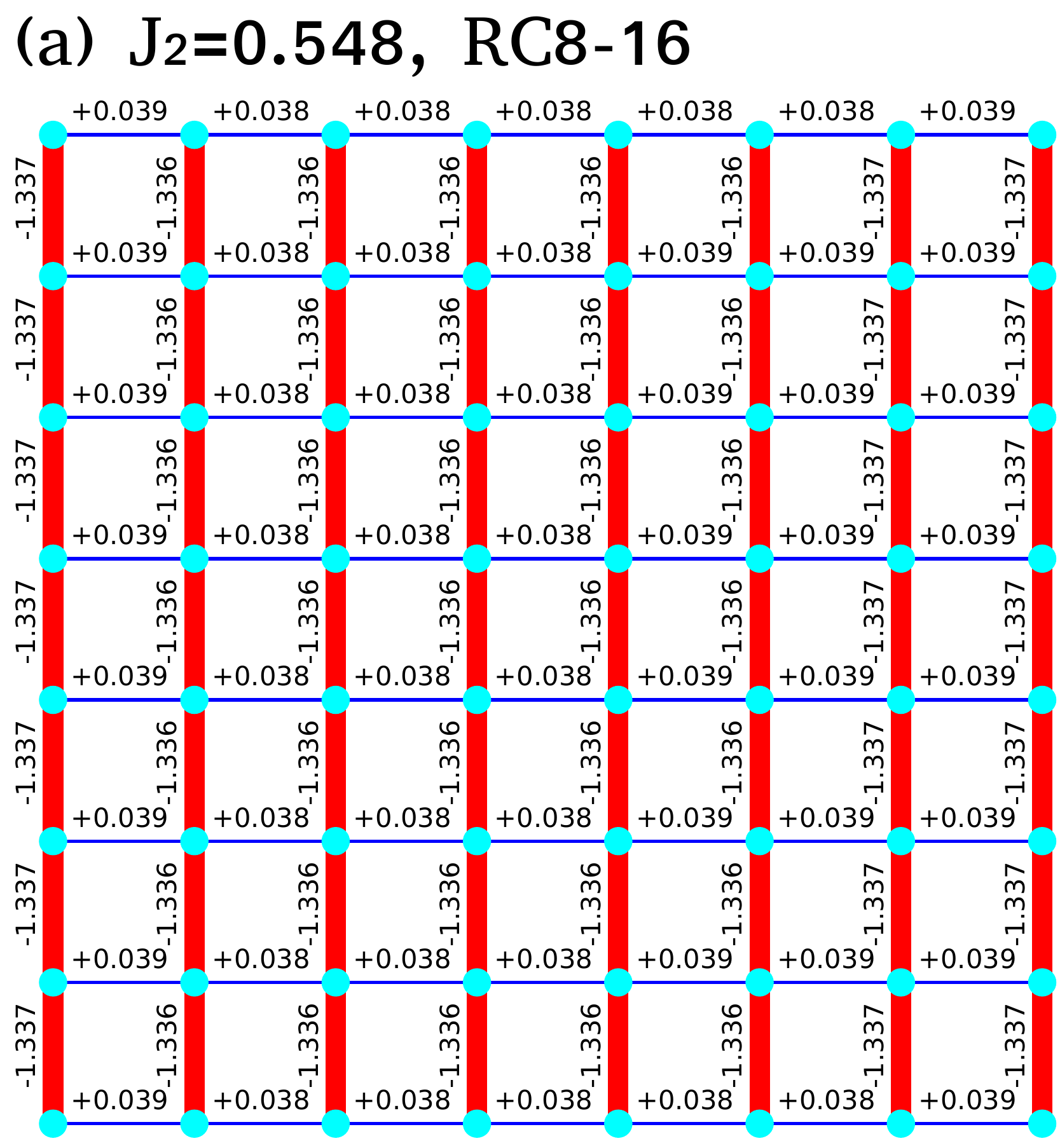}
\includegraphics[width = 0.9\linewidth]{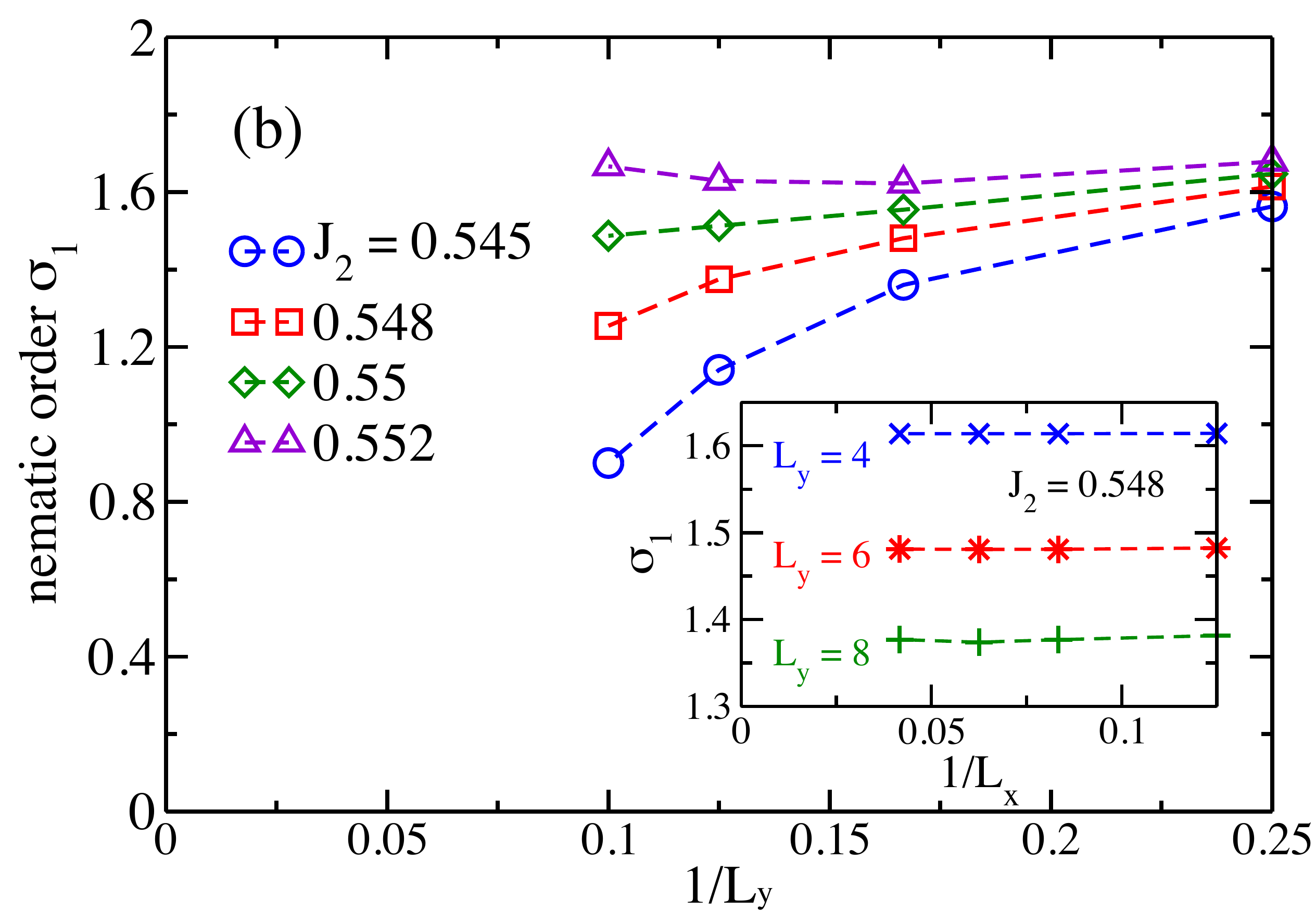}
\caption{(Color online) Lattice orders in the intermediate regime.
(a) Nearest-neighbor $J_1$ bond energy for $J_2 = 0.548$ on the RC8-16 cylinder.
Here we show the middle $8\times 8$ sites. The numbers denote the $J_1$ bond energy
$\langle {\bf S}_i \cdot {\bf S}_j \rangle$. (b) Finite-size scaling of the lattice 
nematic order parameter $\sigma_1$ for different $J_2$ couplings. The inset shows 
the weak dependence of $\sigma_1$ on the cylinder length $L_x$.}
\label{fig:nematic_j1j2}
\end{figure}
%%%%%%

Since the different orders may break different lattice symmetries, here we study the lattice order by calculating the nearest-neighbor bond energy $\langle {\bf S}_i \cdot {\bf S}_j \rangle$. In Fig.~\ref{fig:nematic_j1j2}(a), we show the nearest-neighbor $J_1$ bond energy for $J_2 = 0.548$ on the RC8-16 cylinder, which is in the possible intermediate regime. We can see that although the open boundaries of cylinder break lattice translational symmetry along the $x$-direction, the bond energy in the bulk of cylinder is quite uniform. We also find that the bond energy difference on the open edges decays quite fast to the uniform bulk value (not shown here), indicating a very small boundary effect. Thus, the lattice translational symmetry is preserved in the intermediate regime, which is different from the DMRG results of the spin-$1$ $J_1-J_2$ honeycomb model, in which the system shows a strong tendency to form a plaquette VBS in the intermediate phase~\cite{Gong:2015}.

%%%%%% 
\begin{figure}
\includegraphics[width = 1\linewidth]{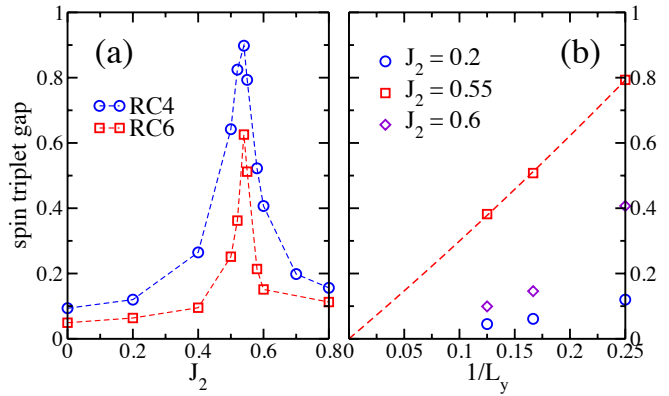}
\caption{(Color online) Spin triplet gap on the finite-size systems.
(a) The triplet gap versus $J_2$ on the $L_y = 4, 6$ cylinders.
(b) Finite-size scaling of the triplet gap versus $1/L_y$ in the different phases.
The spin gap is obtained by sweeping the middle $L_y \times L_y$ sites in the spin-1
sector based on the ground state of the RC$L_y$-$2L_y$ cylinder. For the data at $J_2 = 0.55$,
the gap is fitted by using $1/L_y$ up to the second order.
}
\label{fig:gap}
\end{figure}
%%%%%%

Although the bond energy for $J_2 = 0.548$ in Fig.~\ref{fig:nematic_j1j2}(a) preserves the translational symmetry, it shows a strong bond nematicity. To investigate the possibility of the lattice rotational symmetry breaking, we study the bond nematic order $\sigma_1$, which is defined as $\sigma_1 = \langle {\bf S}_i \cdot {\bf S}_{i+\hat{x}} \rangle - \langle {\bf S}_i \cdot {\bf S}_{i+\hat{y}} \rangle$. For the N\'eel phase with ${\bf q} = (\pi, \pi)$, $\sigma_1$ should be vanished; on the other hand, it should be finite in the stripe phase with ${\bf q} = (0,\pi)/(\pi,0)$. In Fig.~\ref{fig:nematic_j1j2}(b), we show the finite-size scaling of $\sigma_1$ versus $1/L_y$. We should emphasize that although the cylinder geometry has already broken the lattice $C_4$ rotational symmetry, the size scaling has been shown, in different phases, to be an effective way for determining whether the nematic order would be finite or not in the large-size limit~\cite{Hu:2016,Gong:2017}. In the inset, we present the nematic order $\sigma_1$ versus different length $L_x$ with fixed $L_y$. We find that $\sigma_1$ is almost invariant with growing $L_x$, which indicates small finite-size effects along the $x$ direction. Although the results shown in the inset are only for $J_2 = 0.548$, it holds for general $J_2$. For $J_2 \lesssim 0.545$, $\sigma_1$ decays fast and tends to vanish in the thermodynamic limit, which is consistent with the N\'eel phase. For $J_2 \geq 0.55$, $\sigma_1$ goes to a finite value with increasing $L_y$, which agrees with the stripe order breaking lattice $C_4$ symmetry. In the small intermediate regime such as $J_2 = 0.548$, $\sigma_1$ decreases with growing cylinder width; however, because of the system size limit, our DMRG results cannot determine the nematic order in the thermodynamic limit.

We also calculate the spin triplet gap of the system. The triplet gap is defined as $\Delta E = E_1 - E_0$, where $E_1$ is the lowest-energy state in the total spin $S = 1$ sector, and $E_0$ is the ground state energy in the $S = 0$ sector. We calculate the gap by first obtaining the ground state on the RC$L_y$-$2L_y$ cylinder, and then sweeping the middle $L_y \times L_y$ sites in the total spin $S = 1$ sector, avoiding the open edge excitations~\cite{White:2011}. In Fig.~\ref{fig:gap}(a), we demonstrate the triplet gap with growing $J_2$ on the RC4 and RC6 cylinders. The gap grows sharply near $J_2 \simeq 0.55$ on the finite-size systems, which seems to suggest a finite gap. Note that as the convergence challenge for targeting the $S = 1$ sector, we calculate the gap on the $L_y = 8$ cylinder only for a few $J_2$ points. In Fig.~\ref{fig:gap}(b), we show the finite-size scaling of the gap for different $J_2$. For $J_2 = 0.2$ and $0.6$, the fastly decreasing gap is consistent with the magnetic orders spontaneously breaking spin rotational symmetry. For $J_2 = 0.55$, the gap drops fast and seems to be consistent with vanishing in the large-size limit. In our DMRG calculation with the improved system size, the possible intermediate regime shrinks rapidly compared with the previous DMRG result~\cite{Jiang:2009}, which suggests strong finite-size effects and may imply a direct phase transition between the two magnetic order phases that could be consistent with the vanished spin triplet gap.

%%%%%%%%%%%%%%%%%%%%%%%%%%%%%%%%
\section{Summary and discussion}
\label{Sec:CONCLUSION}

We have used the combined numerical methods of the iPEPS ansatz and $SU(2)$ DMRG to study the ground-state phase diagram of the spin-$1$ $J_1-J_2$ Heisenberg model on the square lattice. While the iPEPS ansatz probs the nature of the quantum phase directly in the thermodynamic limit, DMRG obtains accurate results on a cylinder system with finite $L_y$. The final data of the iPEPS and the DMRG calculations are respectively obtained by using the finite-$D$ and finite-$L_{y}$ scaling. 

In our iPEPS simulation, we find that the lowest-energy state transits from the N\'eel state to the stripe state directly at $J_2 \simeq 0.549$. Even if the iPEPS ansatz is biased toward the competitive paramagnetic states (Haldane and VBS), it could not provide lower energy than the magnetically order states. The correlation length near the transition appears to be finite in the large-$D$ limit, supporting a first-order transition between the two magnetic order phases. In the finite-size DMRG calculation, our finite-size scaling analysis up to $L_y = 10$ finds the previous proposed intermediate regime~\cite{Jiang:2009} shrinks rapidly with increasing $L_y$; as such a dramatic change  implies the vanished intermediate phase. In our DMRG results, the stripe order grows up sharply at $J_2 \simeq 0.55$, also supporting a first-order transition consistent with the iPEPS result.

%compared with the previous DMRG results ($0.525 < J_2 < 0.555$) based on the torus geometry up to $L_y = 8$~\cite{Jiang:2009}:
Our study opens up the door to reexamine the phase diagram of the spin-$1$ $J_{1x}-J_{1y}-J_2$ Heisenberg model on the square lattice, where $J_{1x}$ and $J_{1y}$ are the spacial anisotropic nearest-neighbor interactions. While the Schwinger-Boson mean-field theory predicted a fluctuation-induced first-order transition between the N\'eel and stripe phase, which only terminates at a tricritical point for a large anisotropy $(J_{1y} - J_{1x})/J_{1y}$, previous DMRG results suggested a non-magnetic phase emerging near the transition line~\cite{Jiang:2009}. Our results have shown that the non-magnetic phase in the isotropic case ($J_{1x} = J_{1y}$) is unlikely, thus it would be interesting to study whether the anisotropy could enhance quantum fluctuations and open a paramagnetic phase in the intermediate regime.

%%%%%%%%%%%%%%%%
\acknowledgments
The research was partly supported by National Science Foundation Grants PREM DMR-1205734 (R.H.), and by the start-up funding from Beihang University (S.S.G.). Work by DNS was supported by the Department of Energy, Office of Basic Energy Sciences, Division of Materials Sciences and Engineering, under Contract No. DE-AC02-76SF00515 through SLAC National Accelerator Laboratory. We have used \emph{Uni10} \cite{Kao:2015} as a middleware library to build the iPEPS ansatz. 
%%%%%%%%%%%%%%%%%%%%%%
\bibliography{J1-J2}

\begin{thebibliography}{65}
\bibitem [{not()}]{note:hysteresis}%
  \BibitemOpen
  \href@noop {} {}\bibinfo {note} {In case of a N\'eel-to-Stripe first-order
  phase transition, the order state directly cross each other, as at the
  critical point, both equally exist. One expects that slightly above th
  critical point, in the Stripe phase, the N\'eel state would be a pseudo
  ground state: a finite-D iPEPS ansatz initialized by the N\'eel state
  probably fails to converge to the Stripe state. This effect as the order
  states remain metastable around the critical point is referred to a
  hysteresis effect.}\BibitemShut {Stop}%
\end{thebibliography}

\end{document}